\documentclass[twocolumn,prl,preprintnumbers,showpacs,superscriptaddress,%
amsmath]{revtex4}

\usepackage{amsfonts}
\usepackage{hyperref}

\hypersetup{citebordercolor=1 0 0,linkbordercolor=0 0 1,urlbordercolor=0 0 1}

\newcommand{\Zom}{\mathbb{Z}}
\newcommand{\cL}{\mathcal{L}}
\newcommand{\cM}{\mathcal{M}}
\newcommand{\cN}{\mathcal{N}}

\newcommand{\MVM}{\mathcal{M}_\text{VM}}
\newcommand{\MHM}{\mathcal{M}_\text{HM}}

\newcommand{\vb}{\bar{v}}
\newcommand{\I}{\mathrm{i}}
\newcommand{\e}{\mathrm{e}}
\newcommand{\rt}{\mathrm{t}}
\newcommand{\SL}{\mathrm{SL}}
\newcommand{\Li}{\mathrm{Li}}

\renewcommand{\vr}[1]{{\vec{r}\,}^{#1}}

\newcommand{\arch}[1]{[\href{http://arxiv.org/abs/#1}{arXiv:#1}]}

\DeclareMathOperator{\re}{\mathrm{Re}}
\DeclareMathOperator{\im}{\mathrm{Im}}

\begin{document}%%%%%%%%%%%%%%%%%%%%%%%
\preprint{ITP--UU--06/49}
\preprint{SPIN--06/38}
\preprint{YITP--SB--06--53}
\preprint{arXiv:hep-th/0612027}

\title{Nonperturbative corrections to 4D string theory effective actions
from \\ SL(2,Z) duality and supersymmetry}

\author{Daniel Robles-Llana}
\email{D.RoblesLlana@phys.uu.nl}
\affiliation{Institute for Theoretical Physics and Spinoza
Institute, Utrecht University, 3508 TD Utrecht, The Netherlands}

\author{Martin Ro\v{c}ek}
\email{rocek@max2.physics.sunysb.edu}
\affiliation{C.N.~Yang Institute for Theoretical Physics, 
Stony Brook University, NY 11794-3840, USA}

\author{Frank Saueressig}
\email{F.S.Saueressig@phys.uu.nl}
\affiliation{Institute for Theoretical Physics and Spinoza
Institute, Utrecht University, 3508 TD Utrecht, The Netherlands}

\author{Ulrich Theis}
\email{Ulrich.Theis@uni-jena.de}
\affiliation{Institute for Theoretical Physics,
Friedrich-Schiller-University Jena, D--07743 Jena, Germany}

\author{Stefan Vandoren}
\email{S.Vandoren@phys.uu.nl}
\affiliation{Institute for Theoretical Physics and Spinoza
Institute, Utrecht University, 3508 TD Utrecht, The Netherlands}

\begin{abstract}

We find the D($-1$) and D1-brane instanton contributions to the 
hypermultiplet moduli space of type IIB string compactifications
on Calabi--Yau threefolds. These combine with known perturbative and
worldsheet instanton corrections into a single modular invariant
function that determines the hypermultiplet low-energy effective
action.

\end{abstract}

\pacs{11.25.-w, 04.65.+e}

\maketitle %%%%%%%%%%%%%%%%%%%%%%%%%%%%%%%%%%%%%%%%%%%%%%%%%%%%%%%%%%%%%

The absence of a complete nonperturbative formulation of string theory
is its main shortcoming as a full-fledged quantum theory unifying all
known fundamental interactions. Empirically, perturbative weakly coupled
string theory does not give a detailed description of our universe;
consequently, the understanding of nonperturbative phenomena is
essential to making possible a detailed confrontation of string theory
and experiment. Generally, it is difficult to obtain exact information
about nonperturbative structures. However, in special cases such as
those presented here, the symmetries and dualities of the theory are
powerful enough to fix the exact couplings in the low-energy effective
action.

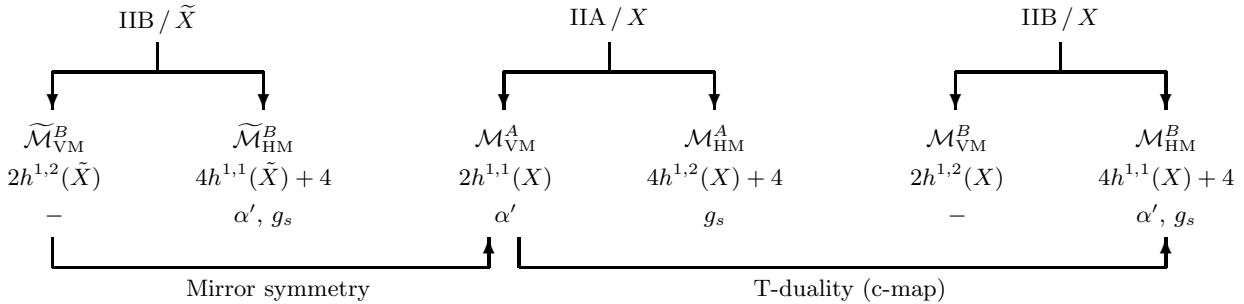
\begin{figure*}[t]
\setlength{\unitlength}{1cm}
\begin{picture}(17.5,4)
\thicklines
\put(2,3.6){\makebox(2,.6){$\mathrm{IIB}\,/\,\widetilde{X}$}}
\put(8,3.6){\makebox(2,.6){$\mathrm{IIA}\,/\,X$}}
\put(14,3.6){\makebox(2,.6){$\mathrm{IIB}\,/\,X$}}
\multiput(3,3.6)(6,0){3}{\line(0,-1){.4}}
\multiput(1.6,3.2)(6,0){3}{\line(1,0){2.8}}
\multiput(1.6,3.2)(6,0){3}{\vector(0,-1){.5}}
\multiput(4.4,3.2)(6,0){3}{\vector(0,-1){.5}}
\put(1,1.2){\shortstack[c]{$\widetilde{\cM}_\mathrm{VM}^B$ \\[1pt]
 $2 h^{1,2}(\tilde{X})$ \\[2pt] $- \vphantom{g'}$}}
\put(3.5,1.2){\shortstack[c]{$\widetilde{\cM}_\mathrm{HM}^B$ \\[1pt]
 $4 h^{1,1}(\tilde{X}) + 4$ \\[2pt] $\alpha'$, $g_s$}}
\put(7,1.2){\shortstack[c]{$\MVM^A$ \\[2pt] $2 h^{1,1}(X)$ \\[2pt]
 $\alpha'$}}
\put(9.5,1.2){\shortstack[c]{$\MHM^A$ \\[2pt] $4 h^{1,2}(X) + 4$ \\[2pt]
 $g_s \vphantom{g'}$}}
\put(13,1.2){\shortstack[c]{$\MVM^B$ \\[2pt] $2 h^{1,2}(X)$ \\[2pt]
 $- \vphantom{\alpha'}$}}
\put(15.5,1.2){\shortstack[c]{$\MHM^B$ \\[2pt] $4 h^{1,1}(X) + 4$ \\[2pt]
 $\alpha'$, $g_s$}}
\put(1.6,1){\line(0,-1){.4}}
\put(7.4,.6){\vector(0,1){.4}}
\put(7.8,1){\line(0,-1){.4}}
\put(16.4,.6){\vector(0,1){.4}}
\put(1.6,.6){\line(1,0){5.8}}
\put(7.8,.6){\line(1,0){8.6}}
\put(3.6,0){\makebox(2,.6){Mirror symmetry}}
\put(11.2,0){\makebox(2,.6){T-duality (c-map)}}
\end{picture}
\parbox[c]{\textwidth}{\caption{\label{eins}{\footnotesize Dualities
relating vector and hypermultiplet moduli spaces arising in CY
compactifications of type II strings. The moduli spaces for vector and
hypermultiplets are denoted by $\MVM$ and $\MHM$, respectively, with
their dimensions below in terms of the Hodge numbers $h^{1,2}$ and
$h^{1,1}$ of the CY $X$ or its mirror $\tilde{X}$ with $h^{1,1}
(\tilde{X})=h^{1,2}(X)$. The lines below indicate the kind of quantum
corrections as well as some of the dualities that relate the various
moduli spaces. Mirror symmetry is the statement that
$\widetilde{\cM}_\text{VM/HM}^B=\cM_\text{VM/HM}^A$, while the c-map
determines $\MHM^{B/A}$ at string tree-level in terms of $\MVM^{A/B}$.}}}
\end{figure*}

The examples we consider are provided by type II string compactifications
on Calabi--Yau threefolds (CY), where the four-dimensional effective
actions are constrained by $\cN=2$ supersymmetry. The massless fields are
components of a supergravity multiplet, vector multiplets, or
hypermultiplets. Their scalar fields parametrize moduli spaces $\MVM$
and $\MHM$, respectively, which locally form a direct product \cite{Nis2}.
The special geometry of $\MVM$ is determined  by a holomorphic function
$F(X)$ \cite{deWit1984}. The exact expression for this function includes
perturbative and worldsheet instanton corrections in the inverse string
tension $\alpha'$, which can in principle be computed by mirror symmetry,
see \emph{e.g.}\ \cite{Hori}. (See Fig.~\ref{eins} for details.)

On the other hand, the string coupling constant $g_s$ is set by the
vacuum expectation value of the dilaton, whose four-dimensional reduction
belongs to a hypermultiplet. Thus $\MHM$ receives both perturbative
and nonperturbative $g_s$ corrections. Building on earlier work
\cite{AFMN,AMTV}, the perturbative corrections have recently been
understood in \cite{RSV}. The nonperturbative corrections arise in the
IIA case from Euclidean D2 or NS5-branes wrapping around supersymmetric
three-cycles or the entire CY\@, respectively, and in the IIB case from
D($-1$)-instantons as well as D1, D3, D5, and NS5-branes wrapping
holomorphic cycles in the CY\@ \cite{Becker1995}. Little is known about
summing up such corrections -- see however \cite{Ooguri1996,Witten1999}
for some results in the limit where gravity decouples.

In this letter, we use the constraints from supersymmetry and the $\SL(2,
\Zom)$ duality symmetry of IIB string theory to determine the full D($-1$)
and D1-brane instanton corrected low-energy effective action for
hypermultiplets in type IIB compactifications on CY. This provides a large
class of four-dimensional $\cN=2$ supergravity theories where exact results
are obtained to all orders in both $\alpha'$ and $g_s$; such results were
not available in four dimensions previously. 

Similar ideas were applied in 
\cite{GG,Green:1997di,Kiritsis1997,Pioline:1997pu} to obtain instanton
corrections to higher dimensional effective actions: D($-1$)
contributions were understood from the modular invariant completion of
the $R^4$ terms in the ten-dimensional effective action, where they are
linked to perturbative $\alpha'$ corrections. The D1-branes belong to
the $\SL(2,\Zom)$ multiplet formed by $(p,q)$-strings
\cite{Schwarz:1995dk,WW}. Therefore their instanton contributions can be
found by applying $\SL(2,\Zom)$ transformations to the worldsheet
instanton contributions, which arise from fundamental strings wrapping
two-cycles of the compact space.

Hitherto, implementing these dualities on the four-dimensional
hypermultiplet moduli space has been hampered by the complicated
quaternion-K\"ahler geometry \cite{BagWit} of $\MHM$. In recent years,
however, it has become clear that, as for the vector multiplet case,
hypermultiplet couplings to $\cN=2$ supergravity are determined by a
single function $\chi(\phi)$ \cite{deWit,deWit2,SdW}. Moreover, when only
D($-1$) and D1-branes are present, enough Peccei--Quinn shift symmetries
in the RR sector remain unbroken so that the hypermultiplet sector can
be described in terms of tensor multiplets. Projective superspace methods
\cite{GHR} and the superconformal tensor calculus \cite{SdW} can then be
used to encode the off-shell tensor multiplet couplings in terms of a
potential $\chi^\rt$. After dualizing tensors into scalars, $\chi^\rt$
coincides with $\chi(\phi)$. This allows us to find the nonperturbative
D($-1$) and D1-instanton corrected low-energy effective action by imposing
the constraints coming from supersymmetry and $\SL(2,\Zom)$ invariance on
$\chi^\rt$. In contrast to the vector multiplet prepotential, the potential
$\chi^\rt$ is real and therefore receives contributions from both
instantons and anti-instantons.

At string tree-level, the c-map \cite{c-map} provides a simple relation
between $\chi^\rt$ and the vector prepotential of the T-dual
compactification on the same CY \cite{RVV}. The IIB hypermultiplet sector
is thus determined by the holomorphic prepotential $F(X)$ of IIA vector
multiplets. This prepotential describes the complexified K\"ahler
deformations of the CY and consists of three parts. The first one encodes
the classical geometry and is given by
 \begin{equation} \label{Fcl}
  F_\text{cl}(X) = \frac{1}{4!}\, \kappa_{abc}\, \frac{X^a X^b X^c}
  {X^1}\ ,
 \end{equation}
where $X^\Lambda=\{X^1,X^a\}$ (with $a$ running over $h^{1,1}$ values)
are the complexified K\"ahler moduli and $\kappa_{abc}$ are the
(classical) triple intersection numbers. This prepotential receives
two kinds of quantum corrections: a perturbative $\alpha'$ correction
proportional to the Euler number $\chi_E=2(h^{1,1}-h^{1,2})$ of the CY\@,
 \begin{equation} \label{Fpt}
  F_\text{pt}(X) = \frac{\I}{8}\, \zeta(3)\, \chi_E\, (X^1)^2\ ,
 \end{equation}
and nonperturbative worldsheet instanton contributions
\begin{equation} \label{Fws}
  F_\text{ws}(X) = -\frac{\I}{4}\, (X^1)^2 \sum_{\{k_a\}} n_{k_a}\,
  \Li_3 \big( \e^{2\pi\I k_a X^a\!/X^1} \big)\ ,
\end{equation}
where the $n_{k_a}$ are genus zero Gopakumar--Vafa invariants \cite{HM,GV}
that enumerate the rational curves of class $k_a$ in the CY (see,
\emph{e.g.}, \cite{Hori}). The polylogarithm appearing here is defined
as $\Li_s(x)=\sum_{n>0}n^{-s}x^n$.

The c-map relates $F(X)$ to the T-dual tensor multiplet Lagrangian density
$\cL$ in superspace \cite{RVV}:
 \begin{equation} \label{RV}
  \cL(v^I, \vb^I, x^I) = \im \oint_C \frac{\mathrm{d}\zeta}{2\pi \I\,
  \zeta}\, \frac{F(\eta^\Lambda)}{\eta^0}\ .
 \end{equation}
Here the $\eta^I\equiv v^I/\zeta+x^I-\vb^I\zeta$, $I=(0,\Lambda)$, are
$\cN=2$ tensor supermultiplets, $\zeta$ is a local coordinate on the
Riemann sphere, and the integration contour $C$ encloses one of the
roots $\zeta_+$ of $\zeta\eta^0$. The Lagrangian \eqref{RV} belongs to
the general class of superconformal tensor multiplet Lagrangians, which
are based on an arbitrary function of the $\eta^I$ homogeneous of degree
one \cite{deWit2}. The corresponding tensor potential (which depends on
the tensor multiplet scalars $v^I,\vb^I,x^I$ only) can be computed via
\cite{SdW}
 \begin{equation} \label{SW}
  \chi^\rt(v^I, \vb^I, x^I) \equiv x^I \partial_{x^I} \cL - \cL\ .
 \end{equation}
In this formulation the physical scalar fields of the Poincar\'e theory
are given by $\mathrm{SU}(2)_R$ and dilatation invariant combinations of
the tensor multiplet scalars. The latter can be grouped into $\mathrm{SU}
(2)_R$ vectors $\vr{I}=(2v^I,2\vb^I,x^I)$ with $\vr{I}\!\cdot\vr{J}=4v^{(I}
\vb^{J)}+x^I x^J$. Taking the superconformal quotient, one obtains the
following identifications \cite{Neitzke:2007ke}: the dilaton-axion system
$\tau=a+\I\e^{-\phi_{10}}=\tau_1+\I\tau_2$ can be expressed as
 \begin{equation} \label{tau} 
  \tau = \frac{1}{(r^0)^2}\, \Big( \vr{0}\! \cdot \vr{1} + \I\, |\vr{0}\!
  \times \vr{1}| \Big)\ ,
 \end{equation}
while the complexified K\"ahler moduli and the remaining RR scalars are
given by \footnote{Alternatively, one may use the type IIA variables
$A^a=\vr{0}\!\cdot\vr{a}/(\vr{0})^2=\tau_1 b^a-c^a$, see
\cite{RSV}.}
 \begin{equation}
  z^a = b^a + \I t^a = \frac{\eta^a(\zeta_+)}{\eta^1(\zeta_+)}\ ,\quad
  c^a = \frac{(\vr{0}\! \times \vr{1})\! \cdot\! (\vr{1}\! \times \vr{a})}
  {(\vr{0}\! \times \vr{1}){\vphantom{\vec{r}}}^2}\ .
 \end{equation}
Substituting \eqref{RV} into \eqref{SW} and expressing the result in terms
of the physical fields, the tensor potential can be computed (after a
numerical rescaling of $r^0=|\vr{0}|$)
 \begin{equation} \label{chi-F}
  \chi^\rt = \sqrt{2}\, r^0 \tau_2^2 \im\! \big( F_1(z) + \bar{z}^a
  F_a(z) \big)\ ,
 \end{equation}
where $F_\Lambda=\partial F/\partial X^\Lambda$. Note that the $r^0$
dependence of $\chi^\rt$ is completely fixed by its superconformal weight.

The tensor potential
 \begin{equation} \label{Psi}
  \chi^\rt = \chi^\rt_\text{cl} + \chi^\rt_\text{pt} + \chi^\rt_\text{ws}
 \end{equation}
describing the tensor multiplet geometry at string tree-level is then
obtained by evaluating \eqref{chi-F} for  $F=F_\text{cl}+F_\text{pt}+
F_\text{ws}$. The classical prepotential \eqref{Fcl} gives rise to
 \begin{equation} \label{chi_class}
  \chi^\rt_\text{cl} = \sqrt{2}\, r^0 \sqrt{\tau_2}\ V(\sqrt{\tau_2}\,
  t)
 \end{equation}
with $V(t)=\tfrac{1}{3!}\kappa_{abc}\,t^a t^b t^c$, the perturbative
$\alpha'$ corrections yield
 \begin{equation} \label{chi_pert}
  2\sqrt{2}\, \chi^\rt_\text{pt} = r^0 \chi_E\, \zeta(3)\, \tau_2^2\ ,
 \end{equation}
and the worldsheet instanton corrections give
 \begin{equation} \label{psiws}
  \sqrt{2}\, \chi^\rt_\text{ws} = - r^0 \tau_2^2 \sum_{\{k_a\}} n_{k_a}
  \re\! \big( \Li_3(\e^{\I w}) + w_2\, \Li_2(\e^{\I w}) \big)
 \end{equation}
with $w\equiv w_1+\I w_2=2\pi k_a z^a$.

We now proceed to implement the $\SL(2,\Zom)$ duality, and then verify
that the resulting nonperturbative completions of \eqref{chi_class} --
\eqref{psiws} indeed satisfy the supersymmetry constraints. The action of
the modular group on the physical scalar and tensor fields has been worked
out in \cite{BGHL}. On the scalar fields, the $S$ generator acts as
 \begin{gather}
  \tau \mapsto -\frac{1}{\tau}\ ,\quad t^a \mapsto |\tau| t^a\ ,\quad
  b^a \mapsto -c^a\ ,\quad c^a \mapsto b^a\ , \label{stransform}
 \end{gather}
while the $T$ transformation shifts
 \begin{equation} \label{ttransform}
  \tau_1 \mapsto \tau_1 + 1\ ,\quad c^a \mapsto c^a + b^a
 \end{equation}
and leaves the other scalars invariant. This action is induced by linear
transformations of the tensor multiplets \cite{B}
 \begin{equation} \label{SL2Zr}
  \eta^0 \mapsto d\, \eta^0 + c\, \eta^1\ ,\quad \eta^1 \mapsto b\,
  \eta^0 + a\, \eta^1\ ,\quad \eta^a \mapsto \eta^a\ ,
 \end{equation}
where the integers $a,b,c,d$ are the entries of the $\SL(2,\Zom)$
transformation matrix and obey $ad-bc=1$. In particular, \eqref{SL2Zr}
implies that under $S$ transformations $r^0\mapsto|\tau|r^0$.

The IIB hypermultiplet effective action should be invariant under the
$\SL(2,\Zom)$ transformation of the four-dimensional fields. In other
words, the $\SL(2,\Zom)$ transformations \eqref{stransform} and
\eqref{ttransform} should act as discrete isometries on the
quaternion-K\"ahler metric underlying the hypermultiplet sector. As
such they should lift to invariances of the (hyperk\"ahler) potential
$\chi(\phi)$ \cite{deWit} and $\chi^\rt$. Note that this condition is
more restrictive than in the K\"ahler case, where invariance of the
metric requires invariance of the K\"ahler potential only up to a
K\"ahler transformation.

Let us examine the modular properties of the tensor potential
\eqref{Psi}. From the transformations \eqref{stransform},
\eqref{ttransform} and \eqref{SL2Zr} it follows that the classical
part \eqref{chi_class} is indeed $\SL(2,\Zom)$ invariant, which is
consistent with the invariance of the classical tensor multiplet action.

The perturbative tensor potential \eqref{chi_pert}, however, is not
modular invariant by itself. The form of its modular invariant completion
becomes apparent after including the perturbative string loop correction
\cite{AMTV,RSV,SdW} 
 \begin{equation} 
  \chi^\rt_\text{1-loop} = \frac{\zeta(2)}{\sqrt{2}}\, \chi_E\, r^0\ .
 \end{equation}
Adding this to \eqref{chi_pert} gives the perturbatively corrected tensor
potential 
 \begin{equation} \label{chi_pert'}
  2\sqrt{2}\, \chi^\rt_\text{pt}{\!\!}' = \chi_E\, r^0 \sqrt{\tau_2}\,
  \Big( \zeta(3)\, \tau_2^{3/2} + 2 \zeta(2)\, \tau_2^{-1/2} \Big)\ .
 \end{equation}
Inside the brackets one recognizes the first two terms in the expansion
of the nonholomorphic Eisenstein series
 \begin{equation} \label{E3/2}
  Z_{3/2}(\tau,\bar{\tau}) = \sideset{}{'} \sum_{m,n} \frac{\tau_2^{3/2}}
  {|m\tau + n|^3}\ ,
 \end{equation}
where the primed sum is taken over all integers $(m,n)\in\Zom^2
\backslash(0,0)$. This series is naturally generated by applying the
method of images to \eqref{chi_pert'}. Because the prefactor $r^0
\sqrt{\tau_2}$ is $\SL(2,\Zom)$ invariant, the modular invariant
completion of \eqref{chi_pert'} is
 \begin{equation} \label{Psi_-1}
  2\sqrt{2}\, \chi^\rt_{(-1)} = \frac{\chi_E}{2}\, r^0 \sqrt{\tau_2}\,
  Z_{3/2}(\tau,\bar{\tau})\ .
 \end{equation}
This expression encodes the perturbative and the full nonperturbative
D($-1$)-brane instanton corrections to the four-dimensional hypermultiplet
effective action. After Poisson resummation (with $K_1$ the modified
Bessel function of first order)
 \begin{align}
  Z_{3/2} & = 2 \zeta(3)\, \tau_2^{3/2} + \frac{2\pi^2}{3}\,
  \tau_2^{-1/2} + 8\pi\, \tau_2^{1/2}\, \times \notag \\
  & \quad\, \times\!\!\! \sum_{m\neq 0,n>0} \Big| \frac{m}{n} \Big|\,
  \e^{2\pi \I mn \tau_1}\, K_1(2\pi |mn| \tau_2)\ ,
 \end{align}
we find terms that are exponentially suppressed at large $\tau_2$; these
have a clear interpretation as D($-1$)-brane instanton contributions. This
is analogous to the ten-dimensional IIB case, where the perturbative and
D($-1$)-brane corrections combine into $\sqrt{\tau_2}\,Z_{3/2}$ to give
the modular invariant and supersymmetry preserving pre\-factor of the
$R^4$ terms (in string frame) \cite{GG,Green:1997di}.

As for the perturbative correction in \eqref{chi_pert} discussed above,
the nonperturbative term $\chi^\rt_\text{ws}$ is not $\SL(2,\Zom)$
invariant. This was to be expected, since $\chi^\rt_\text{ws}$ contains
the contributions of the fundamental string wrapping supersymmetric cycles
of the CY, whereas the other contributions from $(p,q)$-strings in the same
$\SL(2,\Zom)$ multiplet are missing. These can be taken into account by
summing $\chi^\rt_\text{ws}$ over all its $\SL(2,\Zom)$ transforms, modulo
the stabilizer group $\Gamma_{\!\infty}$ generated by $T$. The result is
 \begin{align} \label{chi_inst}
  2\sqrt{2}\, \chi^\rt_{(1)} = - r^0 \sqrt{\tau_2}\, & \sum_{\{k_a\}}
  n_{k_a}\sideset{}{'} \sum_{m,n} \frac{\tau_2^{3/2}}{|m\tau + n|^3}\,
  \times \notag \\
  & \times \big( 1 + |m\tau + n| w_2 \big)\, \e^{-S_{m,n}}
 \end{align}
with instanton action
 \begin{equation}
  S_{m,n} = 2\pi k_a \big( |m\tau + n|\, t^a - \I m\, c^a - \I n\, b^a
  \big)\ .
 \end{equation}
The potential \eqref{chi_inst} encompasses all instanton corrections due
to wrapped F- and D-strings as well as their bound states. Indeed,
\eqref{stransform} and \eqref{ttransform} imply
 \begin{equation}
  T:\ S_{m,n} \mapsto S_{m,n+m}\ ,\qquad S:\ S_{m,n} \mapsto S_{-n,m}\ ,
 \end{equation}
which are precisely the $\SL(2,\Zom)$ transformations of the
$(p,q)$-string action \cite{WW}. Note that the instanton corrections
\eqref{chi_inst} are completely fixed by the topology of the underlying
CY through its Gopakumar--Vafa invariants.

Finally, we must verify that the modular invariant tensor potentials
constructed above satisfy the constraints arising from supersymmetry
\cite{SdW}. The key observation is that the supersymmetry constraints,
when formulated at the level of $\chi^\rt$, transform covariantly under
global general linear transformations of the tensor multiplets $\eta^{I}
\mapsto{\Lambda^I}_J\,\eta^{J}$, with $\Lambda\in\mathrm{GL}(n,
\mathbb{R})$. In particular, the $\SL(2,\Zom)$ transformations
\eqref{SL2Zr} are embedded in this group, and hence commute with
supersymmetry. Thus, if we start with the supersymmetric potentials
$\chi^\rt_\text{pt}$ and $\chi^\rt_\text{ws}$ and add their $\SL(2,\Zom)$
transforms, the sum is also supersymmetric. Consequently, the modular
invariant tensor potential
 \begin{equation}
  \chi^\rt = \chi^\rt_\text{cl} + \chi^\rt_{(-1)} + \chi^\rt_{(1)}
 \end{equation}
leads to a supersymmetric tensor multiplet Lagrangian.

We conclude with some remarks. A crucial ingredient in our derivation
has been the off-shell tensor multiplet formulation of the low-energy
effective supergravity action. After dualizing the tensors into scalars,
one obtains the nonperturbative corrections to the hypermultiplet moduli
space, together with additional isometries acting as continuous shifts on
the dual scalars. These shift symmetries will however be broken by
three-brane and five-brane instantons, which have not been included in
our work. The main reason is that this sector transforms into itself under
$\SL(2,\Zom)$: D3-branes are self\-dual, and D5-branes are mapped
into NS5-branes and vice versa. Hence, they cannot be generated from the
D($-1$), D1-brane and worldsheet instantons with the method used in this
letter. To determine these corrections as well, a better understanding of
the full duality group beyond $\SL(2,\Zom)$ is needed. This U-duality
group acts nontrivially on the hypermultiplet moduli space and will
transform all instanton corrections into each other. Implementing the
full symmetry group will require new methods beyond the ones used in this
paper (see \cite{ARV} for a speculation on how some such corrections might
be described). We leave this as an open problem, whose solution will
undoubtedly contribute to a better understanding of nonperturbative
string theory.

\begin{acknowledgments} %%%%%%%%%%%%%%%%%%%%%%%%%%%%%%%%%%%%%%%%%%%%%%%%

This work was initiated at the 4th Simons Workshop in Physics and
Mathematics. DRL\@, FS\@, UT and SV thank the YITP and the Department of
Mathematics at Stony Brook University for hospitality and partial
support.\ DRL is supported by the European Union RTN network
MRTN-CT-2004-005104.\ MR is supported in part by NSF grant no.\
PHY-0354776.\ FS is supported by the European Commission Marie Curie
Fellowship no.\ MEIF-CT-2005-023966.

\end{acknowledgments}

\raggedright

\end{document}